
\magnification=1200
\vsize=9.0 true in
\hsize=6.5 true in
\voffset=.55 true in
\tolerance=1200
\parindent=24pt
\parskip=07pt
\baselineskip=13pt
\rm

\centerline{\bf A Status Report on Chiral Dynamics: Theory and Experiment}
\vskip .20 true in
\centerline{Aron M. Bernstein}
\centerline{Laboratory for Nuclear Science and Department of Physics}
\centerline{Massachusetts Institute of Technology, Cambridge, MA \ 02139}
\vskip .15 true in
\centerline{Barry R. Holstein}
\centerline{Department of Physics and Astronomy}
\centerline{University of Massachusetts, Amherst, MA \ 01003}
\vskip .50 true in
\baselineskip=24pt
\centerline{\bf Abstract}
\vskip .15 true in

This paper gives an overview of a Workshop on Chiral Dynamics: Theory and
Experiment which was held at MIT July 25-29, 1994; its unique feature was
the equal mixture of theory and experiment.  The purpose of the workshop
was to bring together many of the active participants to assess the status
of the field and to explore fruitful future directions.  The foundations
and present status of the theory were reviewed as well as the experimental
status of the field.  To facilitate the discussion of future directions
working groups were organized on Threshold Photo/Electropion Production,
Two-Pion Threshold and the Chiral Anomaly, Nucleon Polarizabilities, Pion
and Sigma Polarizabilities, $\pi -\pi$ Interactions, and $\pi$-N
Interactions.
\vfill\eject

The main attraction of chiral dynamics [Weinberg, Leutwyler,
We79,Ga83]{\baselineskip=12pt \footnote*{References to talks at the
workshop shall be denoted by the name of the author in parenthesis.  The
summary of the workshop will be published by Springer-Verlag as {\it Chiral
Dynamics: Theory and Experiment}, A.M. Bernstein and B.R. Holstein,
editors.  Whenever possible these will be supplemented by references to the
published literature.}} is
that it represents a rigorous and model-independent methodology by which to
make QCD predictions at the confinement scale.  This is a forefront
arena of the standard model and as such it is essential to confront these
predictions with precise experiments.  Interest in this field has been
growing rapidly due to increasing confidence in the theory and also as a
result of a new generation of accelerators and experimental techniques.
Chiral dynamics, or chiral perturbation theory, is an (low energy)
effective  field theory for  QCD. At the confinement scale where the
interaction between the quarks is very strong the relevant degrees of
freedom are the baryons ${\rm N, \Sigma , \Delta}$, {\it etc}.  and the
Goldstone bosons ${\rm \pi , \eta , K.}$ A  perturbative treatment is
appropriate since the Goldstone interactions are relatively weak at low
energies. Indeed the Goldstone theorem requires the vanishing of all
interactions of Goldstone bosons at zero energy-momentum.  The effective
field theory  obeys all of the underlying symmetries of the QCD Lagrangian.
Of particular relevance is chiral symmetry, which is broken both
dynamically and also explicitly by the small but non-zero light quark
masses $({\rm m_u, m_d, m_s})$.

Isospin is another important approximate symmetry which was
discussed by both Weinberg and Leutwyler.  At a superficial
level one might expect large isospin  violations due to the
rather large ratio ${\rm  r = (m_d - m_u)/ (m_d + m_u) \sim
0.3}$.  However, the observed  isospin violation is much
smaller due to the fact that the up and down quark masses
are  small compared to the QCD  scale and the relevant
isospin breaking quantity is ${\rm (m_d - m_u)/\Lambda_{QCD}
\sim  0.01}$ [Leutwyler], which is of the same order as
electromagnetic effects.  Nevertheless there are some special
cases which involve the s-wave $\pi^0$ nucleon scattering or
charge exchange reactions (in which the $\pi^0$ appears in either the
final and/or initial states) for which this isospin violation can be larger
[We79], and this possibility, which is the focus of a
new experimental initiative, is discussed below.

The price that must be paid for dealing with an effective,
nonrenormalizable theory is the appearance of a number of
unknown low energy constants, which must be determined experimentally.
In principle they can also be obtained from the QCD Lagrangian by
integrating out the high energy degrees of freedom, and it
has been shown that their magnitudes can be reliably estimated by
saturating with the low lying resonances [Ecker,Ec89] or very
approximately determined by lattice gauge calculations [Negele,My94]. A
related approach, which appears promising, is based on a chiral
version of the Schwinger-Dyson equation [Roberts,Ro94].  In this model
an ansatz must be made for the gluon distribution function.
However, after that is done observables can be calculated without
any additional parameters.

The predictions of chiral perturbation theory are expressed
as a power series in energy-momentum and in current quark mass
with a scale parameter ${\rm \sim  4 \pi \ F_\pi \sim \ 1 \ GeV}$. The
convergence properties of the theory are determined by the structure of
the process under consideration and in  a scattering problem, for example,
will depend on the position of the lowest resonances and on the thresholds
for particle production. The most accurately predicted quantities are
the properties of the Goldstone bosons and their strong and electroweak
interactions. In this  sector there is a unique connection between the
number of loops and the corresponding power of the energy (momentum).
Although most existing results are one loop or  ${\cal O}(p^4)$, in a few
cases two loop---${\cal O}(p^6)$---calculations are becoming available.

In the relativistic treatment of the ${\rm \pi N}$ sector the
simple relationship between the number of loops and the
power of energy (momentum) is not valid due to the presence
of the nucleon mass as an additional parameter with the
dimensions of energy. However, in the heavy
fermion version of the theory the simple
power counting is restored. There has been impressive
progress in this arena in the past few years, as discussed
by Mei\ss ner and Bernard[Be95a].

{}From the previous discussion one can observe a sort of
``complementarity'' about the  experimental tests of chiral
dynamics. On the one hand it is the Goldstone boson sector
of the theory which is most amenable to reliable
calculations. However, since one cannot make targets of
these unstable particles more difficult, indirect methods,
must be employed in order to confront these predictions with
experimental tests.  The most mature example, summarized by
Po\v cani\'c, is the low energy $\pi-\pi$ interaction which
was first predicted using current algebra and then refined
with chiral perturbation theory. The original experimental
technique used the nucleon as a source of virtual pions,
with the $\pi-\pi$   interaction  determined via
extrapolation to the pole at ${\rm t = m_\pi^2}$ (here t is
the invariant four momentum transfer, which is negative in
the physical region, hence the extrapolation).  An alternative
approach involves
measurements of the final state $\pi-\pi$ interaction in the
near-threshold ${\rm \pi  N \rightarrow \pi \pi  N}$
reaction, for which there exists a significant data base.  The largest
error at the present time in the latter method is associated with
the phenomenological
method used in order to extract the $\pi-\pi$ scattering lengths from
the data.  The $\pi-\pi$ interaction has also been measured
in the ${\rm  K^+ \rightarrow  \pi^+ \pi^-e^+\nu_e}$  decay
(${\rm  K_{e4}}$ decay, branching ratio $\sim 3.9 \times
10^{-5}$).  At the present time there appears to exist a
discrepancy between results obtained via different methods
[Po\v cani\' c], as shown in Figure 1.
However, due to possible systematic errors
in the extraction of the scattering lengths from the
${\rm \pi N \rightarrow \pi \pi N}$ reaction and the low
statistics of the ${\rm K_{e4}}$ decay data, it is premature
to draw  definitive conclusions. At the present level of
precision,  it is clear that more needs to be done.
We can look forward to improved theoretical analysis of the
${\rm \pi N \rightarrow \pi \pi N}$ reaction and to  vastly
improved ${\rm K_{e4}}$ decay data from the Frascati
$\Phi$ factory ${\rm DA\Phi NE}$ [Gasser and Sevior, Baldini
and Pasqualucci, Da92]. In addition, there are plans to measure
the $\pi-\pi$  scattering length by observation of pionium
(bound $\pi^+ \pi^-$  atoms) in experiments at CERN and Indiana.

An additional area in which we can look forward to new
experimental results is the study of the chiral anomaly. In
this case parameter-free predictions can be made for
threshold reactions. An anomaly is said to occur for a
situation in which a symmetry of the classical
Lagrangian is not obeyed when the theory is quantized.
The most famous  and successful example in particle physics
is the prediction for the $\pi^0 \rightarrow \gamma \gamma$
decay rate. We can anticipate new results for the $\gamma
\pi \rightarrow \pi \pi$ reaction, for which there also
exists a solid prediction based on the chiral anomaly.  Although this
prediction strictly speaking obtains only at the unphysical center
of mass energy $\sqrt{s}=0$, it can with reasonable assumptions be
extended into the physical domain.  There exists an approved
experiment at CEBAF to study this reaction using virtual pions from
a proton target in the reaction $ep\rightarrow \pi^+\pi^0ne'$
as well as at Fermilab, which will utilize a high energy pion
beam and the virtual, {\it i.e. Primakoff effect}, photons
from a high Z nuclear target [Miskimen, Moinester].

Hadron polarizabilities $\bar{\alpha}$ and $\bar{\beta}$
(electric and magnetic), which measure the response of systems to the
presence of external electromagnetic fields, are important probes of
internal
structure.  Again the ``complementarity'' between theory and
experiment applies in that the predictions for the (hard
to measure) pion are considerably more precise than for the
nucleon, for which abundant targets are available.
The nucleon situation was summarized by Nathan. For
the proton, which  has been studied for over 30 years, there
has been considerable recent progress[Ha93]. At the present time
the most precise results are obtained using the forward
scattering (unitarity) sum rule constraint for
$\bar{\alpha} + \bar{\beta}$ while fitting  $\bar{\alpha} -
\bar{\beta}$  from the Compton scattering measurements.  For
the neutron the only precise data has been obtained from
elastic neutron scattering from Pb combined with a
sum rule constraint[Sc91]. The results are
compared to theoretical predictions in Table 1. The
${\cal O}(p^4)$ chiral perturbation theoretical results
[Mei\ss ner] are shown with errors which
take into account uncertainties associated with  the
relevant low energy constants. It can be seen that for
$\bar{\beta}$ (the magnetic polarizability), where the
contribution of the $\Delta$ is significant, the theoretical
uncertainties are correspondingly large. Within errors,
however,  there is reasonable agreement between theory and
experiment.  Although this is encouraging, it is clear that
more precise data, particularly on the neutron, is
highly desirable. It is also very important to clarify the
relationship between the formulations of baryon chiral
perturbation theory with and without the $\Delta$ as an
explicit participant, and to reduce the theoretical
uncertainty [Butler and Nathan].
\vfill\eject
\centerline{{\bf Table I.} Nucleon Polarizabilities (${\rm
10^{-43} \ fm^3)}$}
\vskip -.50 true in
$$\hbox{\vbox{\offinterlineskip\def\strut{\hbox{\vrule
height 10pt depth 2pt width 0pt}}
\halign{\strut#\hfil&\tabskip 0.4 true in &
#\hfil&
#\hfil  \tabskip 0.0 in \cr
\noalign{\hrule\vskip .05 true in}

& Experimental$^\dagger$ & Theory$^\ast$ \cr
\noalign{\hrule\vskip .25 true in}

$\bar{\alpha}_p$ & $12.0 \pm 0.9$  & $10.5 \pm 2.0$ \cr
\noalign{\vskip .125 true in}

$\bar{\alpha}_n$ & $12.5 \pm 2.50$ & $13.4 \pm 1.5$ \cr
\noalign{\vskip .125 true in}

$\bar{\beta}_p$ & $\phantom{1}2.2 \pm 0.9$ & $\phantom{1}3.5
\pm 3.6$ \cr
\noalign{\vskip .125 true in}

$\bar{\beta}_n$ & $\phantom{1}3.3 \pm 2.7$ & $\phantom{1}7.8
\pm 3.6$ \cr
\noalign{\vskip .125 true in}

$\bar{\alpha}_p + \bar{\beta}_p$ & $14.2 \pm 0.5$ & $14.0
\pm 2.2$ \cr
\noalign{\vskip .125 true in}

$\bar{\alpha}_n + \bar{\beta}_n$\phantom{Theory} & $15.8 \pm
1.0 $ & $21.1 \pm 2.2$ \cr
\noalign{\vskip .125 true in \hrule}
\cr}}}$$
\vskip -.60 true in
\hskip 1.25 true in $^\ast$CHPT [Mei\ss ner] \hskip .25 true in
$^\dagger$[Nathan, Ha93,Sc91]

In the case of pion polarizabilities, much attention has been focussed upon
the $\gamma\gamma\rightarrow\pi^0\pi^0,\pi^+\pi^-$, channels which
will be studied at DA$\Phi$NE.  In the $2\pi^0$ case, discrepancies between
SLAC experimental measurements and  a one loop chiral perturbation theory
calculation have recently been resolved via a dispersion relation estimate
as well as a full two loop calculation, as shown in Figure 2[Gasser,Be94a].
A similar two-loop calculation is underway for the corresponding
$\pi^+\pi^-$ process.  However, in this case the one-loop chiral and
dispersive calculations are already in good agreement with each other and
with experiment.  At the present time there exist several experimental
measurements of the pion polarizabilities, which are, however, highly
divergent.  The working group on hadron polarizabilities [Baldini and
Bellucci] discussed three active areas for future measurements. These are:
1) the ${\rm \pi^+ Z \rightarrow  \gamma  \pi^+ Z}$  reaction
from virtual photons in a high Z target (Primakoff effect) using high
energy pions at Fermilab (${\rm K^+ \ and \ \Sigma^+}$  polarizabilities
will be measured in a similar fashion) [Moinester]; 2) the ${\rm \gamma p
\rightarrow \gamma \pi^+ n}$ reaction (radiative photoproduction)
extrapolated to the pion pole planned at Mainz; and 3) the ${\rm e^+ e^-
\rightarrow  e^+ e^- \pi^+ \pi^-  (\pi^0 \pi^0)}$  reaction planned at
${\rm DA\Phi NE}$ [Baldini and Pasqualucci]. It has been demonstrated that
the latter reaction is an excellent probe of the dynamics but is
unfortunately relatively insensitive to the pion polarizabilities [Gasser,
Baldini and Bellucci, Do93]. We can look forward to results from each of
these experimental initiatives in the next few years.

The  ${\rm \pi N}$ interaction at low energies is a fundamental testing
ground for any theory of the strong interactions. Of particular interest
for chiral dynamics is the sigma term  which is predicted to be  ${\rm
\sigma  = (35\pm 5)}$ MeV/(1 - y) where y is a measure of the strange quark
content of the nucleon [Sainio,Ga91]. There has been a long history of
determining $\sigma$  from the data which was reviewed by Sainio and
H\"ohler. The complications include 1) an extrapolation must be made from
the physical region to the Cheng-Dashen point by an analytic continuation
of the empirically determined partial wave scattering amplitudes;
2) the relevant amplitude is isoscalar which is relatively small; 3) there
exist systematic discrepancies in the data base; and 4) there is a rapidly
varying t-dependence of the scalar form factor of the nucleon, which is
determined by dispersion relations (lowest order chiral perturbation
theory calculations are insufficient). To illustrate the problems
associated with point 2 and 3 above, the sensitivity of the differential
cross section to the sigma term for the elastic scattering of 30 MeV pions
from the proton is shown in Figure 3[Sa94].  It can be seen that
the variation of the cross section, for a range in the sigma term
corresponding approximately to the present uncertainty in its
determination, is about the same  as the systemic error between the data
sets. Despite all of these difficulties the value ${\rm y = 0.2 \pm 0.2}$,
corresponding to a contribution of 130 MeV to the nucleon mass, has been
determined [Sainio,Ga91]. It is clearly desirable to  reduce the 100\%
uncertainty in y. Of related interest is the suggestion by Leutwyler to
measure the $\pi\pi$ sigma term from the observation of the $\pi-\pi$
scattering lengths in the pionium atom.

The interesting possibility of observing isospin violation
due to the mass difference of the up and down quarks was
discussed by Weinberg and  Leutwyler. There was also a
review of the present status of isospin violation in the
$\Delta$ region [H\"ohler] as well as the presentation of
several recent developments. In particular, new
measurements of the ${\rm  \pi^- p \ and \ \pi^- p
\rightarrow \pi^0 n}$ charge exchange scattering lengths
have been performed at PSI by observing the transition
energy and width of the ${\rm 3p \rightarrow 1s}$ transition
in pionic hydrogen [A. Badertscher {\it et al} . in the $\pi
N$ working group summary]. The values obtained show a
possible violation of isospin symmetry [Achenauer {\it et
al} . in the $\pi N$ working group summary].  There is also
a proposed new technique by which to measure  ${\rm \pi^+ n
\rightarrow  \pi^0 p}$ and possibly ${\rm \pi^0 p}$
scattering lengths in the ${\rm \gamma p \rightarrow \pi^0
p}$ reaction as a final state interaction effect [Be94]. All
of the scattering lengths depend on the value of the ${\rm
\pi N}$ coupling constant as was discussed by Pavan [${\rm
\pi N}$ working group summary] and H\"ohler.

Threshold electromagnetic pion production is an additional
testing ground for chiral dynamics [Bernard, Mei\ss ner and
Schoch,Be95b]. The field dates back to current algebra derivations
of ``low energy theorems'' for threshold pion photoproduction. However, it
has only been in the past five years that accurate experimental data have
been available. A flurry of activity was spurred by experiments at Saclay
and Mainz on the threshold ${\rm  p(\gamma , \pi^0)}$ reaction [Ma86,Be90].
When the dust settled down it was found that these experiments agreed with
the soft pion predictions [Be91]. However, it was subsequently realized
that the ``low energy theorems'' omitted important loop contributions and
at the workshop Ecker demoted them to ``low energy guesses.'' It now
appears that the situation is more complicated with a substantial fraction
of the prediction for the threshold electric dipole amplitude ${\rm
E_{0+}}$  coming from an undetermined low energy constant which must be
determined from experiment [Bernard,Be95b]. Its value cannot be predicted
with precision at the present time. Despite this situation, a {\it new} low
energy theorem has been derived for two of the three p wave threshold
multipoles [Bernard,Be95b], and these predictions agree with the
Mainz data. However, since an unpolarized cross section does not uniquely
determine the multipoles this does not constitute a proof of these
predictions are correct. In fact there is another empirical solution for
the Mainz data which disagrees with the new p wave numbers [Be94]. A
definitive determination of the multipoles awaits the results of more
recent experiments  for the  (unpolarized) cross sections performed at
Mainz [St90] and SAL [Bergstrom] as well as an experiment planned with
linearly polarized photons at Mainz.

It was pointed out by Bernard that once the relevant low
energy constants are determined by the data for the
${\rm p(\gamma , \pi^0)}$ reaction, there are no additional
free parameters in the predictions for the corresponding
${\rm p( e, e' \pi^0)}$ process.  It is therefore of
increasing importance to make precision measurements of the
threshold electroproduction reaction. Two preliminary
results were reported at the workshop [Blok, Walcher,Va95] from
NIKHEF and Mainz for ${\rm q^2 = -0.1 \ GeV^2}$  and for
CM energies W up to 15 MeV above threshold.  At Mainz the longitudinal and
transverse contributions to the cross sections have been obtained
(Rosenbluth separation).  The results of this analysis will be available
soon.

As the magnitude of ${\rm q^2}$  increases so does the relative
contribution of the one loop contribution, reaching 50\% of the total cross
section by q$^2$=-0.1 GeV$^2$.  Therefore it is suggested that precision
measurements of the ${\rm p(e,e'\pi^0)}$ reaction be carried out at a
smaller magnitude of q$^2$ [Bernard].  An experiment at ${\rm q^2 = -.05 \
GeV^2}$ is planned at Mainz during the next year.

The previous discussion has been focused on neutral pion production. For
charged pions one expects the old low energy theorems to be more accurate
since the leading (Kroll-Ruderman) term for the  threshold electric dipole
amplitude ${\rm E_{0+}}$ comes from a marriage of gauge  and chiral
invariance  and is model-independent. Nevertheless the existing
experimental tests are based on older emulsion measurements and
involve relatively large extrapolations to threshold.
There are plans to perform the ${\rm p(\gamma ,
\pi^+)n}$ reaction at Saskatoon [Bergstrom]. Also
preliminary results for the ${\rm p(\pi^-, \gamma)n}$
reaction near threshold were presented by Kovash for a
TRIUMF  experiment. Finally we note the emergence of the
$(\gamma,2\pi)$ reaction as an interesting object of study of
study [Bernard, Walcher].  Here preliminary measurements in
the $\pi^0\pi^0$ channel by the TAPS detector at Mainz have revealed
a rather substantial signal in the near-threshold region.

The entire discussion of the ${\rm \pi N}$ sector is based
on the physics of light (almost massless) up and down
quarks. In the purely mesonic sector it is customary to
treat the {\it three} light quarks (u, d, s). This SU(3)
treatment has yet to be extended to the meson-baryon
sector which would then include ${\rm \pi , \eta, K}$ as
well as nucleons, hyperons (and $\Delta$,$\Sigma^*$ ?).
Measurements in this extended sector are already in
progress at Bates where the SAMPLE experiment is looking for
the strange quark content of the nucleon [Kowalski].
Likewise experiments at Bonn and Mainz and planned
experiments at CEBAF are investigating kaon and eta photo-
and electro-production. In the mesonic sector we also look
forward to exciting new data on the $\pi-\pi$ interaction
from ${\rm K_{e4}}$ decays and from the two photon
production of pion pairs at DA$\Phi$NE [Baldini and
Pasqualucci].

Although, in the interest of brevity, we have not discussed
all of the new experimental and theoretical initiatives that
are applicable to QCD studies at the confinement scale we
hope that we have conveyed some of the sense of  excitement
and  promise that was exhibited at the workshop. We
anxiously look forward to future progress in this field.

\vskip .40 true in
\leftline{\bf References}

\vskip 0pt
\advance\leftskip by .20 true in
\item{[Al82]} E.A. Alekseeva  {\it et al}., Zh. Eksp. Teor. Fiz {\bf 82},
1007
(1982) [Sov. J. JETP, {\bf 55}, 591 (1982)].

\item{[Be90]} R. Beck {\it et al}., Phys. Rev. Lett. {\bf
65}, 1841 (1990).

\item{[Be91]} A. M. Bernstein and B. R. Holstein, Comments
Nucl. Part. Phys. {\bf 20}, 197 (1991), J. Bergstrom, Phys.
Rev. {\bf C44}, 1768 (1991), D. Dreschsel and L. Tiator, J.
Phys. G: Nucl. Part. Phys. {\bf 18}, 449 (1992).

\item{[Be94]} A. M. Bernstein, to be published, and $\pi$N Newsletter No.
9, December 1993, Proceedings of the 5th International Symposium on Meson-
Nucleon Physics and the Structure of the Nucleon, Boulder, Colorado,
September 1993, G. H\" ohler, W. Kluge, and B. M. K. Nefkens, editors.

\item{[Be94a]} S. Bellucci, J. Gasser and M.E. Sainio, Nucl. Phys.
{\bf B423}, 80 (1994).

\item{[Be95a]} V. Bernard, N. Kaiser and Ulf-G. Mei\ss ner, hep-ph9501384,
to be published in Int. J. Mod. Phys. E.

\item{[Be95b]} V. Bernard, N. Kaiser and Ulf-G. Mei\ss ner, to be
published and Nucl. Phys. {\bf B383}, 442 (1992).

\item{[Da92]} DA$\Phi$NE Physics Handbook, Ed. L. Maiani, G. Pancheri and
N. Paver (INFN, Frascati, 1992).

\item{[Do93]} J.F. Donoghue and B.R. Holstein, Phys. Rev. {\bf D48}, 137
(1993).

\item{[Ec89]} G. Ecker, J. Gasser, A. Pich and E. deRafael, Nucl.
Phys. {\bf B321}, 311 (1989); J.F. Donoghue, C. Ramirez and G.
Valencia, Phys. Rev. {\bf D39}, 1947 (1989).

\item{[Ga83]}  J. Gasser and H. Leutwyler, Phys. Lett. {\bf
125B}, 321 (1983); Ann. Phys. N.Y. {\bf 158}, 142 (1984);
Nucl. Phys. {\bf 250B}, 465, 517, 539 (1985).

\item{[Ga91]} J. Gasser, H. Leutwyler and M.E. Sainio, Phys. Lett. {\bf
B253}, 252 and 260 (1991);

\item{[Ha93]} E. Hallin {\it et al}., Phys. Rev. {\bf C48}, 1497 (1993);
B.J. MacGibbon, University of Illinois Ph.D. Thesis (1994), unpublished.

\item{[Ma86]} E. Mazzucato {\it et al}., Phys. Rev. Lett.
{\bf 57}, 3144 (1986).

\item{[MA90]} H. Marsiske et al., Phys. Rev. {\bf D41}, 3324 (1990).

\item{[My94]} S. Myint and C. Rebbi, Nucl. Phys. {\bf B421}, 241 (1994).

\item{[Pe92]} M. Pennington in DA$\Phi$NE Physics Handbook, Ed. L. Maiani,
G. Pancheri and N. Paver (INFN, Frascati, 1992).

\item{[Po94]} D. Pocanic {\it et al}., Phys. Rev. Lett. {\bf 72}, 1156
(1994);
H. Burkhardt and J. Lowe, Phys. Rev. Lett. {\bf 67}, 2622 (1991).

\item{[Ro77]} L. Rosselet {\it et al}., Phys. Rev. {\bf D15}, 574 (1977);
M.M. Nagels {\it et al}., Nucl. Phys. {\bf B147}, 189 (1979).

\item{[Ro94]} C.D. Roberts and A.G. Williams: {\it Dyson-Schwinger
Equations and their Application to Hadronic Physics}, in {\it Progress in
Particle and Nuclear Physics}, Vol. 33, pp. 477-575, ed. by A. Fa\ss ler
(Pergamon Press, Oxford, 1994).

\item{[Sa94]} M. Sainio, private communication (1994).

\item{[Sc91]} J. Schmeidmeyer {\it et al}., Phys. Rev. Lett.
{\bf 66}, 1015 (1991).

\item{[St90]} Mainz experiment A2/7-90, TAPS collaboration,
H. Str\"oher spokesman.

\item{[Va95]} H. B. van den Brink {\it et al}., Phys. Rev. Lett. {\bf 74},
3561 (1995).

\item{[We79]}  S. Weinberg, Physica {\bf 96A}, 327 (1979).
\vskip 0pt
\advance\leftskip by -.20 true in
\vskip 0pt
\vfill\eject

\noindent
{\bf Figure 1}:  The I=2 vs. I=0 s-wave the $\pi\pi$ scattering length
predictions (symbols) and experimental results (contour limits).  The
``soft pion'' results are from the near threshold $\pi N-\pi\pi N$
reaction[Po94]; the ``Chew-Low'' results are from the peripheral $\pi N-
\pi\pi N$ reaction[Al82]; and the ``$K_{e4}$+Ror'' results are from the
$K_{e4}$ decay data using the constraint of the Roy equations[R077].
Dashed line: Weinberg's constraint. Calculations:  Weinberg (full square),
Schwinger (filled triangle),Chang and Gursey (filled inverted triangle),
Jacob and Scadron (open circle), Gasser and Leutwyler-ChPT (open square),
Ivanov and Troitskaya-QLAD (open triangle), Lohse {\it et al}.-Meson
Exchange (open rhomb), Ruivo {\it et al}. and Bernard {\it et al}.-NJL
(open stars), Kuramashi {\it et al}.-quenched lattice gauge QCD (open
cross), Roberts {\it et al}.-GCM (filled stars).
\vskip .25 true in
\noindent
{\bf Figure 2}: The $\gamma\gamma\rightarrow\pi^0\pi^0$ cross section
$\sigma(|\cos\theta|\leq Z)$ as a function of the center of mass
energy E at Z=0.8, together with the data from the Crystal Ball
experiment[Ma90].  The solid line is the full two-loop result, and the
dashed line results from the one-loop calculation[Gasser,Be94a].  The band
denoted by the dash-dotted lines is the result of a dispersive
calculation by Pennington[Pe92].
\vskip .25 true in
\noindent
{\bf Figure 3}: Differential cross section for $\pi^+p$ elastic scattering
at $p_\pi=97$ MeV/c (30 MeV pion kinetic energy).  The curves show the
variation of the predicted cross sections for a variation of $\Sigma$ of
10 MeV which is the range of uncertainty in the $\pi N$ sigma term[Sa94].
The points show the experimental data.

\end